\documentclass{aa}  

\usepackage{lscape}
\usepackage{graphicx}
\usepackage{txfonts}
\usepackage{amsmath}
\usepackage{multirow}
\usepackage{adjustbox}
\usepackage{tabularx}
\usepackage{amsmath}
\usepackage{bm}
\usepackage{array}
\usepackage{xcolor}

\urldef{\webecf}\url{https://erosita.mpe.mpg.de/dr1/eSASS4DR1/eSASS4DR1_arfrmf/}

\newcommand{\eros}{eROSITA\;}

\begin{document}

   \title{The eROSITA Upper Limits:}

   \subtitle{Description and access to the data}

   \author{Dusán Tubín-Arenas\thanks{\email{dtubin@aip.de}}
          \inst{1,2}
          \and
         Mirko Krumpe\inst{1}
         \and
         Georg Lamer\inst{1}
         \and
          Jonas Haase\inst{3}
          \and
          Jeremy Sanders\inst{3}
         \and
         Hermann Brunner\inst{3}
          \and
          David Homan\inst{1}
          \and
          Axel Schwope\inst{1}
          \and
          Antonis Georgakakis\inst{4}
          \and
          Katja Poppenhaeger\inst{1}
          \and
          Iris Traulsen\inst{1}
          \and
          Ole K\"onig\inst{5}
          \and
          Andrea Merloni\inst{3}
          \and
          Alain Gueguen\inst{3}
          \and
          Andrew Strong\inst{3}
          \and
          Zhu Liu\inst{3}
         }

   \institute{Leibniz-Institut f\"ur Astrophysik Potsdam (AIP), An der Sternwarte 16, 14482 Potsdam, Germany
         \and
    Potsdam University, Institute for Physics and Astronomy, Karl-Liebknecht-Straße 24/25, 14476 Potsdam, Germany 
\and
        Max-Planck-Institut f\"ur extraterrestrische Physik, Gießenbachstraße 1, 85748 Garching, Germany
        \and
        Institute for Astronomy \& Astrophysics, National Observatory of Athens, V. Paulou \& I. Metaxa, 11532 Athens, Greece
        \and 
        Dr. Karl Remeis-Observatory and Erlangen Centre for Astroparticle Physics, Friedrich-Alexander-Universit\"at Erlangen-N\"urnberg, Sternwartstr. 7, 96049 Bamberg, Germany}

   \date{Received ...; accepted ...}

 
  \abstract
   {The soft X-ray instrument eROSITA on board the Spectrum-Roentgen-Gamma (SRG) observatory has successfully completed four of the eight planned all-sky surveys, detecting almost one million X-ray sources during the first survey (eRASS1). 
   The catalog of this survey will be released as part of the first eROSITA data release (DR1).  
   }
   {Based on X-ray aperture photometry, we provide flux upper limits for eRASS1 in several energy bands. We cover galactic longitudes between $180^{\circ}\lesssim l \lesssim 360^{\circ}$ (eROSITA-DE). These data are crucial for studying the X-ray properties of variable and transient objects, as well as non-detected sources in the eROSITA all-sky survey data. }
   {We performed aperture photometry on every pixel of the SRG/eROSITA standard pipeline data products for all available sky tiles in the single detection band ($0.2 - 2.3$ keV). Simultaneously, we performed the same analysis in the three-band detection at soft ($0.2-0.6$ keV), medium ($0.6-2.3$ keV), and hard ($2.3-5.0$ keV) energy bands. Based on the combination of products for the individual bands, we are also able to provide aperture photometry products and flux upper limits for the $0.2 - 5.0$ keV energy band. The upper limits were calculated based on a Bayesian approach that utilizes detected counts and background within the circular aperture.}
   {The final data products consist of tables with the aperture photometry products (detected counts, background counts, and exposure time), a close-neighbor flag, and the upper flux limit based on an absorbed power-law spectral model ($\Gamma=2.0, \; N_{\rm H}=3\times10^{20}$ cm$^{-2}$). The upper limits are calculated using the one-sided $3\sigma$ confidence interval (CL) of a normal distribution, representing CL = 99.87\%.
   The aperture photometry products allow for an easy computation of upper limits at any other confidence interval and spectral model. These data are stored in a database with hierarchical indexes in order to offer a fast query option. 
   }
   {We provide a detailed description of the process of retrieving SRG/eROSITA upper limits for a large set of input positions, as well as of the eROSITA data, the X-ray aperture photometry, the upper limit calculation, and the final data products. The eROSITA upper flux limits represent either the maximum flux of potential non-detections or the $3\sigma$ upper flux uncertainty of detected sources. We emphasize the importance of choosing the right spectral model that ought to match the spectral shape of the source of interest: the wrong spectral model can produce discrepancies of up to $30\%$ in the final flux upper limit value. We also describe the architecture of the database and the web tool, which are designed to handle large queries on input positions.}

\authorrunning{D. Tubín-Arenas et al.}

   \maketitle
%

\section{Introduction}

X-ray emission provides valuable insights into a broad variety of physical and cosmological processes across cosmic time and distances. At large scales, high energy emission allows the characterization of the hot plasma that is gravitationally bound inside the potential well of large dark matter halos. This emission thereby traces the mass of the dark matter content, the growth of large-scale structures, and the properties of the intra-cluster medium (ICM) \citep[see, e.g.,][]{2002ARA&A..40..539R,2011ARA&A..49..409A,2017A&A...606A.118H}. Bright X-ray emission at the center of galaxies tracks the accretion of matter onto supermassive black holes (SMBH) which is linked to galaxy formation and galaxy evolution \citep[see, e.g.,][]{2005Natur.433..604D,2008ApJS..175..356H,2012ARA&A..50..455F}. In the local universe, X-ray emitters comprise objects such as single or binary stars, white dwarfs, cataclysmic variables, isolated neutron stars, and stellar mass black holes. These sources are powered by various physical processes such as accretion, thermonuclear explosions, magnetic field decay, or stellar spin-down. 

The extended ROentgen Survey with an Imaging Telescope Array  \citep[eROSITA;][]{2021A&A...647A...1P} is the soft X-ray instrument on board the {\it Spectrum-Roentgen-Gamma} \citep[SRG;][]{2021A&A...656A.132S} observatory. It combines a large field of view (FoV $\sim 1^{\circ}$), effective scanning observation mode, and high sensitivity in the X-ray band ($0.2 - 2.3$ keV), making it the most efficient imaging survey telescope in the X-ray \citep[][]{2012arXiv1209.3114M,2021A&A...647A...1P}. 

Following its launch in 2019 and in order to test the sensitivity, image quality, and spectroscopic capabilities of eROSITA, a calibration and performance verification (CalPV) phase was performed before starting the planned four years of all-sky scanning observations. The most relevant results are presented in the eROSITA Final Equatorial Depth Survey \citep[eFEDS; see][for references and catalog description]{2022A&A...661A...1B}. Altogether, eROSITA observed an area of 140 deg$^{2}$ for 360 ks, resulting in a primary catalog of 27 910 X-ray sources detected in the $0.2-2.3$ keV energy range with a flux limit of $6.5 \times 10^{-15}\; \rm erg\; cm^{-2}\; s^{-1}$.
Using 11 eROSITA CalPV observations, \cite{2022A&A...664A.126L} presented a serendipitous source catalog with 9515 X-ray sources. 

Since the beginning of the eROSITA all-sky survey \citep[eRASS;][]{2021A&A...647A...1P} in December 2019, eROSITA has successfully completed four of the planned eight all-sky scans. In order to complete a single all-sky scan, eROSITA follows a survey strategy with a ``scan rate'' of 0.025 deg s$^{-1}$ and a ``survey rate'' of 1 deg per day. The scan rate consists of a revolution around the observing axis every 4 hr (referred to as ``eROday'') while covering each sky position for $\sim$40 s in the FOV. The survey rate describes the progression of the overlapping eROdays, which cover the whole sky in approximately 182 days (half a year). Following this strategy, eROSITA detected $\sim 1$ million sources during its first scan and it is expected that the final stacked eROSITA all-sky survey will be 25 times more sensitive than the {\it ROSAT} all-sky survey \citep[RASS,][]{1999A&A...349..389V}. The all-sky data products have been organized into 4700 sky tiles that are overlapping square regions of $3.6^{\circ} \times 3.6^{\circ}$ in size and stored in FITS \citep{1981A&AS...44..363W} files for a given energy range. Since eROSITA is a Russian-German collaboration, the sky is equally split into two hemispheres over which each team has unique scientific data exploitation rights. These data rights are separated by Galactic longitude ($l$) and latitude ($b$), with a division marked by the great circle passing through the Galactic poles $(l,b)=(0,+90);(0,-90)$ and the Galactic Center Sgr\,A* $(l,b)=(359.94423568,-0.04616002)$: data with $-0.05576432< l <179.94423568$ degrees (Eastern Galactic hemisphere) belong to the Russian consortium, while data with $359.94423568 > l >179.94423568$ degrees (western galactic hemisphere) belong to the German eROSITA consortium (eROSITA-DE).

The eROSITA source-detection process, described in detail in \cite{2022A&A...661A...1B}, is performed over the $0.2-2.3$ keV band. This single band covers the most sensitive energy range given by the shape of the eROSITA response \citep{2021A&A...647A...1P}. Simultaneously, an independent three-band detection is executed in the soft ($0.2-0.6$ keV), medium ($0.6-2.3$ keV), and hard ($2.3-5.0$ keV) energy bands. According to simulations \citep{2022A&A...661A..27L}, the three-band detection is used to select sources with very soft or hard spectral shapes and create an independent hard-band-selected catalog. 

Despite the invaluable legacy of the eROSITA all-sky survey and the millions of cataloged X-ray sources, many non-detected sources are hidden in the noise of the observations. These non-detections can be faint X-ray objects or intrinsically variable X-ray sources \citep[see, e.g.,][]{Ptak_1998,2004PThPS.155..170U,2004astro.ph.10551V,2006ARA&A..44...49R,2012A&A...544A..80G,2014A&A...563A..57S,2017A&A...603A.127S,2019NewAR..8501524I,2020arXiv201009005D,2021MNRAS.505.1954Z}, for which the emission in their quiescent states can fall below the threshold of a formal detection. However, eROSITA can still constrain their X-ray emission based on the fact that these objects have not been detected in the observations. This limit of how much flux is allowed in an observation without detecting a source is referred to as the "flux upper limit." 
A large number of sources that are detected at wavelengths other than X-rays need X-ray flux upper limits to understand their physical and statistical properties, even if those sources are not formally detected in X-rays. Thus, the X-ray eROSITA upper limits can contribute to important scientific goals such as the investigation of long-term X-ray variability, the search for transients, timing analysis, or X-ray properties of non-X-ray selected sources.

There are several ways to compute upper limits for non-detected sources \citep[see e.g.,][]{1986ApJ...303..336G,1991ApJ...374..344K,vanDyk_2001,2004ApJ...608..957A,2007ApJ...657.1026W}. \cite{2010ApJ...719..900K} provide a detailed theoretical framework related to upper limit and sensitivity calculations and how to interpret those values. We note that \cite{2010ApJ...719..900K} defined the ``upper limits'' as the maximum flux that a source can have without exceeding the detection threshold and the ``upper bound'' as the largest value of the flux inference range of any source. Nowadays, the X-ray community usually refers to the former as ``sensitivity,'' which characterizes the detection process and depends only on the background level, the exposure time, and the detection threshold. We use the term ``upper limit'' to indicate the upper edge of the confidence interval that is computed from the probability distribution of the observed counts at the position of interest. The upper limit is then computed based on the observed counts, background level, and probability distribution. Thus, the upper limits are independent of the detection process and therefore, different from the sensitivity at a certain sky position (see \S \ref{sec:theory}). 

In practice, recent X-ray upper limit projects have been focused on the implementation of web platforms that derive and provide the aforementioned upper flux limits for several X-ray missions. For example, the uninterrupted operations of ESA's {\it XMM-Newton} \citep{2001A&A...365L...1J} mission during the pointed and the Slew Survey \citep{2008A&A...480..611S} observations have provided a sizable archival database with hundreds of thousands of cataloged sources (4XMM catalog: \citealt{2020A&A...641A.136W} and XMM stacked observations: \citealt{2020A&A...641A.137T}). In order to characterize non-detected positions on the sky and obtain upper limits on the {\it XMM-Newton} data, \cite{2022MNRAS.511.4265R} created the RapidXMM database which provides pre-computed upper limit values based on the aperture photometry in the {\it XMM-Newton} pointed and Slew Survey footprint. Another platform that provides X-ray upper limits is the High-Energy Light-curve Generator \citep[HILIGT;][]{SAXTON2022100531,KONIG2022100529} that was implemented to deliver the long-term light curves of a source based on X-ray data from past and current missions with ESA contributions. HILIGT enables users to query and fully exploit the X-ray history of a source for up to 50 years. HILIGT provides a framework that returns the upper flux limit based on aperture photometry of the science images of the corresponding missions or, optionally, a catalog entry if a source was detected within the given aperture. The Living {\it Swift}-XRT Point Source Catalogue \citep[LSXPS;][]{10.1093/mnras/stac2937} and the real-time transient detector is designed to perform low-latency searches of transients and to provide upper limits within the {\it Swift}-XRT footprint.

In this context, we aim to compute and provide eROSITA upper limits of the first eROSITA survey, in the German half of the sky. Our server will deliver photometric products such as detected counts, background counts, exposure times, and close-to-source flags for every pixel of the eROSITA data at the single band detection ($0.2-2.3$ keV). We also provide upper limits and photometric products for the three-band detection run at soft ($0.2-0.6$ keV), medium ($0.6-2.3$ keV), and hard ($2.3-5.0$ keV) energy bands, and for the energy range between $0.2-5.0$ keV.

The paper is organized as follows: In \S \ref{sec:theory}, we describe the theoretical background behind the Bayesian approach used to calculate upper limits. We characterize in \S \ref{sec:calculation} the relevant eROSITA standard pipeline data products and then we describe the algorithm that computes and produces the final upper limit products. In \S \ref{sec:ecf}, we describe the recipe to obtain flux upper limits at any preferred spectral model. The access to the upper limit data via download or via web tool is described in \S \ref{sec:webtool}. Finally, we give our conclusions in \S \ref{sec:conclusions}. 
This work uses data that are public together with the first eROSITA public data release \citep[DR1,][]{Merloni2024}. These data were processed with the eROSITA standard processing pipeline version c010, which is based on the eROSITA Standard Analysis Software System \citep[eSASS,][]{2022A&A...661A...1B}. See \citet{Merloni2024} for a description of the catalogs, the available products, and processing pipeline versions.
Any additional analysis of eROSITA data for this work was performed 
with eSASS version eSASSusers\_211214.
Future upper-limit products, including individual bands and the stacked data of the subsequent eROSITA all-sky survey, will be published based on future pipeline versions.

\section{Theoretical background}\label{sec:theory}

Our calculation of the upper limit of a non-detected source relies on X-ray aperture photometry and it follows the Bayesian approach described by \cite{1991ApJ...374..344K}. We start with a description of the theoretical background behind the upper limit calculation.

The number of detected X-ray photons at a given position and within a certain aperture in the observation is Poisson-+distributed and the probability distribution is given by

\begin{equation}
    P(N\;|\;S + B) = \frac{(S+B)^{N} \cdot e^{-(S+B)}}{N!},
\end{equation}\label{poisson}

\noindent where $P(N\;|\;S + B)$ is the probability of observing $N$ counts, given the expected number of counts $(S+B)$ in the aperture. The expectation value consists of $S$, which is the expected number of counts coming from the source of interest, and $B,$ which takes into account the expected number of counts coming from the background. A typical X-ray observation always provides the total number of counts, $N$, while we assume that the expected background counts, $B,$ can be obtained at a high level of precision in source-free regions, thus neglecting any error associated with $B$. Since we aim to study the source contribution $S$ to the observed number of photons $N$ inside the aperture, we make use of Bayes' Theorem\footnote{The definition of Bayes' Theorem is: $P(A\;|\;B)=\frac{P(B\;|\;A)\cdot P(A)}{P(B)}$, where $P(A\;|\;B)$ is the posterior probability function for parameter A. $P(B\;|\;A)$ is the conditional probability that the event B is true given the event A. Finally, $P(A)$ and $P(B)$ are referred to as the prior and marginal probabilities, respectively.} to obtain the continuous posterior probability function for the parameter $S$ as a function of the observables $N$ and $B$. Thus, we have

\begin{equation}
\label{post}
    P(S\;|\;N,B) = C \cdot \frac{(S+B)^{N} \cdot e^{-(S+B)}}{N!}
,\end{equation}

\noindent the posterior distribution $P(S\;|\;N,B)$ gives the probability that the total observed number of counts ($N$) could have been produced by a source with $S$ counts \citep{1991ApJ...374..344K}. The conditional distribution follows the Poisson distribution. Then, $C$ takes into account the prior knowledge of $S$ and, since the integral of $P(S\;|\;N,B)$ over all $S$ is not normalized, $C$ also works as the normalization factor of the posterior distribution. Assuming non-negativity in $S$ and a uniform prior distribution, $C$ is derived from the normalization requirements as:

\begin{equation}
\label{prior}
\begin{split}
    C^{-1}={} & \int_{0}^{\infty} \frac{(S+B)^{N} \cdot e^{-(S+B)}}{N!} dS \\
={} &\int_{B}^{\infty} \frac{T^{N}\cdot e^{-T}}{N!} dT = \Gamma(N+1,B).
\end{split}
\end{equation}

\noindent Changing the integration variable to $T=S+B$ in Eq.~\ref{prior}, $C$ can be expressed as an upper\footnote{Upper incomplete gamma function: $\Gamma(a,x)= \int_{x}^{\infty} t^{a-1} e^{-t} dt$} incomplete gamma function $\Gamma(N+1,B)$. 

\begin{figure}
  \resizebox{\hsize}{!}{\includegraphics{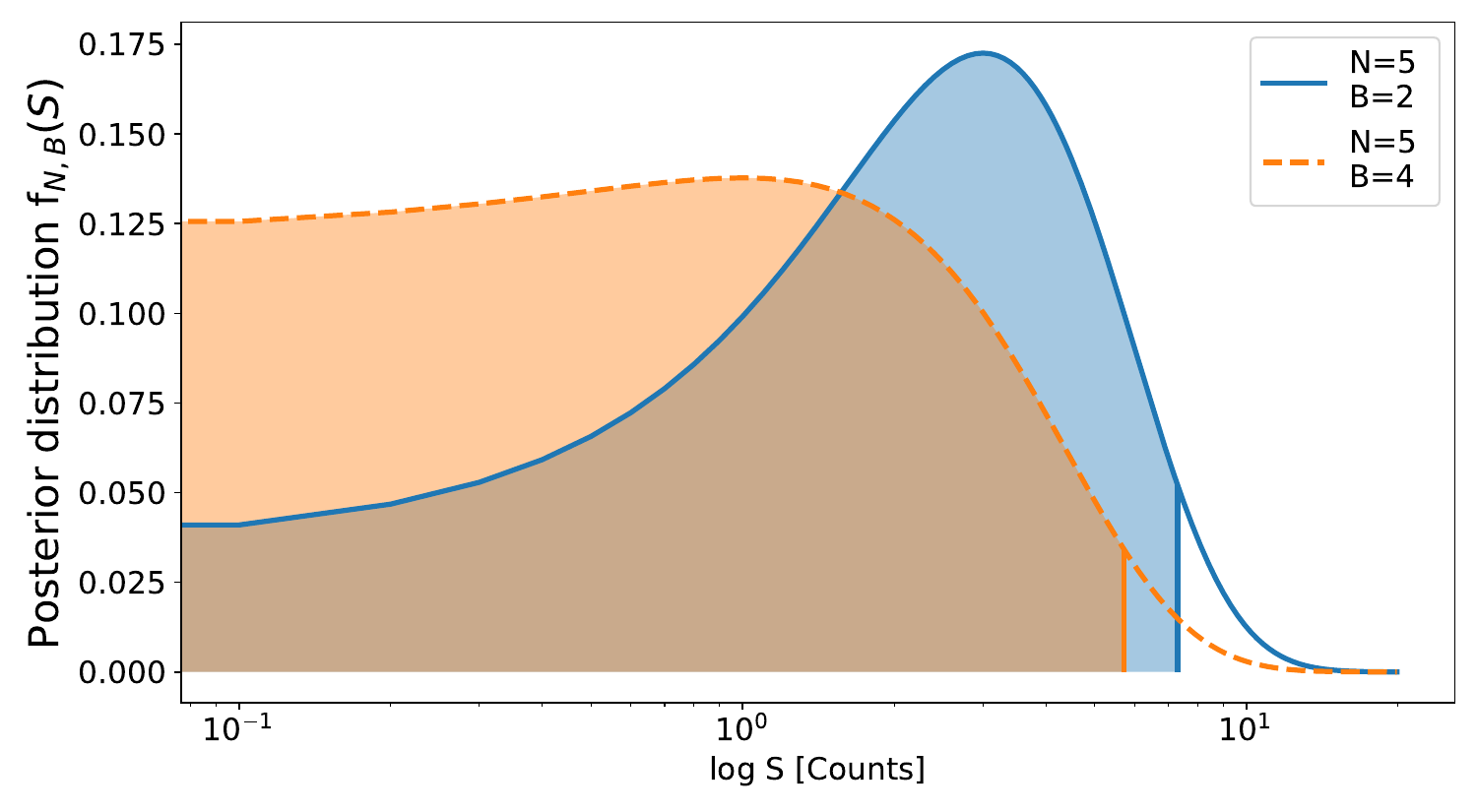}}
  \caption{Posterior distribution function from Eq.~\ref{post} as a function of $\log S$ with a confidence interval of $CL=0.9$ ($90\%$). Both curves correspond to the posterior distribution of an X-ray observation with $N=5$ observed counts but different background levels of $B=2$ (blue solid curve) and $B=4$ counts (orange dashed curve). The shaded area under each curve illustrates the cumulative distribution from Eq.~\ref{cumulative} up to $90\%$ of the cumulative probability. The vertical lines correspond to the one-sided upper limit (UL) calculated from Eq.~\ref{ul}. We find $UL_{5,2}=7.30$ and $UL_{5,4}=5.72$ counts. }
  \label{pic:posterior}
\end{figure}

Finally, it is possible to define the upper limit (UL), in units of counts, as the number of counts $S$ at which the cumulative probability of the posterior distribution in Eq. \ref{post} equals a certain confidence level, $CL$:

\begin{equation}
\label{cumulative}
     C \int_{0}^{UL}  \frac{(S+B)^{N} \cdot e^{-(S+B)}}{N!} dS = CL
.\end{equation}

\noindent Here, $CL$ takes fractional values between 0 and 1 and represents the confidence that the true value of counts coming from the source is contained in the confidence interval, namely, between 0 and the upper limit $UL$.
Using the aforementioned change of the integration variable, \cite{2022MNRAS.511.4265R} derived the following expression for the upper limit: 

\begin{equation}
\label{ul}
    UL = \gamma^{-1}(N+1,CL\cdot\Gamma(N+1,B)+\gamma(N+1,B)) -B. 
\end{equation}

We note that the $UL$ only depends on the number of observed counts, $N$, the expected background level, $B$, and the input confidence level. The upper-limit $UL$ is expressed in terms of the lower\footnote{Lower incomplete gamma function: $\gamma(a,x)= \int_{0}^{x} t^{a-1} e^{-t} dt$} $\gamma(a,x)$ and upper $\Gamma(a,x)$ incomplete gamma functions. 

Figure \ref{pic:posterior} shows the posterior distribution of an X-ray observation with $N=5$ counts and two different background levels of $B=2$ (blue solid curve) and $B=4$ (orange dashed curve). We chose a single-sided confidence interval of $CL = 0.9$ (or $90\%$) represented by the shaded area under each curve. The vertical lines represent the upper limits UL for both distributions and indicate that the true value of photons from the source $S$ lies below this limit, with a confidence of 90\%. We note that for an observation with a fixed number of observed counts, the upper-limit UL is lower when the background level is higher. This is because the probability distribution function, $P(S\;|\;N,B),$ has a single local maximum at $S = N-B $, which is the most probable value for the counts emitted by the source, $S$. This reflects the fact that the higher the background, the less room there is for counts coming from the source. 

In the case of a region where $N\gg B$ (likely a detected source), the probability distribution function can be numerically integrated in both directions starting from the most probable value $S = N - B $ until the confidence interval is reached. Thus, although it is not the main purpose of the paper, it is possible to derive asymmetric lower and upper limits for the counts extracted inside the aperture \citep{1991ApJ...374..344K}. 

Equation \ref{ul} provides the upper limit in units of counts. The upper count rate limit ($CR_{UL}$), in units of counts per second, is written as

\begin{equation}
\label{eq:cr}
    CR_{UL} = \frac{UL}{t \cdot EEF},
\end{equation}

\noindent where $t$ is the exposure time and EEF is the encircled energy fraction; this is the fraction of the point-spread function (PSF) used to define the radius of the extraction aperture, which is included to correct for the source photons that fall outside the aperture.

Finally, source flux upper limits are defined as
\begin{equation}
\label{eq:fl}
    f_{X} = \frac{UL}{t\cdot EEF\cdot ECF}
,\end{equation}
\noindent where ECF is the energy-to-count conversion factor. The ECF depends on the instrument and the spectral shape of the source (e.g., a power law with a given photon index or black-body models). We explain in \S \ref{sec:ecf} how to derive the ECF for any spectral model using XSPEC.

Following the method set out above, an upper flux limit can be derived for any X-ray observation purely based on aperture photometry products, even if the hypothetical source is too faint to be detected.

\section{eROSITA upper limit calculation}\label{sec:calculation}
The theoretical background of \S \ref{sec:theory} presents the principal equations required to compute upper flux limits for a given X-ray observation. In this section, we describe the required initial eROSITA standard pipeline data products, the data extraction that leads to the upper-limit calculation, and the description of the final eROSITA upper-limit products. 
In the following, we focus our analysis on the description of the aperture photometry and upper limit calculations performed over the single energy band $0.2-2.3$ keV. The figures will be based on the data from the sky tile 174069. However, we tested our procedures with several sky tiles at different energy bands. We also produced the upper limits of the three-band detection at individual energy ranges covering the soft ($0.2-0.6$ keV), medium ($0.6-2.3$ keV), hard ($2.3-5.0$ keV), and the summed three-band ($0.2-5.0$ keV), as described in \S \ref{sec:ecf}.

\subsection{Initial eROSITA data}\label{erodata}

\begin{figure*}
   \resizebox{\textwidth}{!}
            {\includegraphics{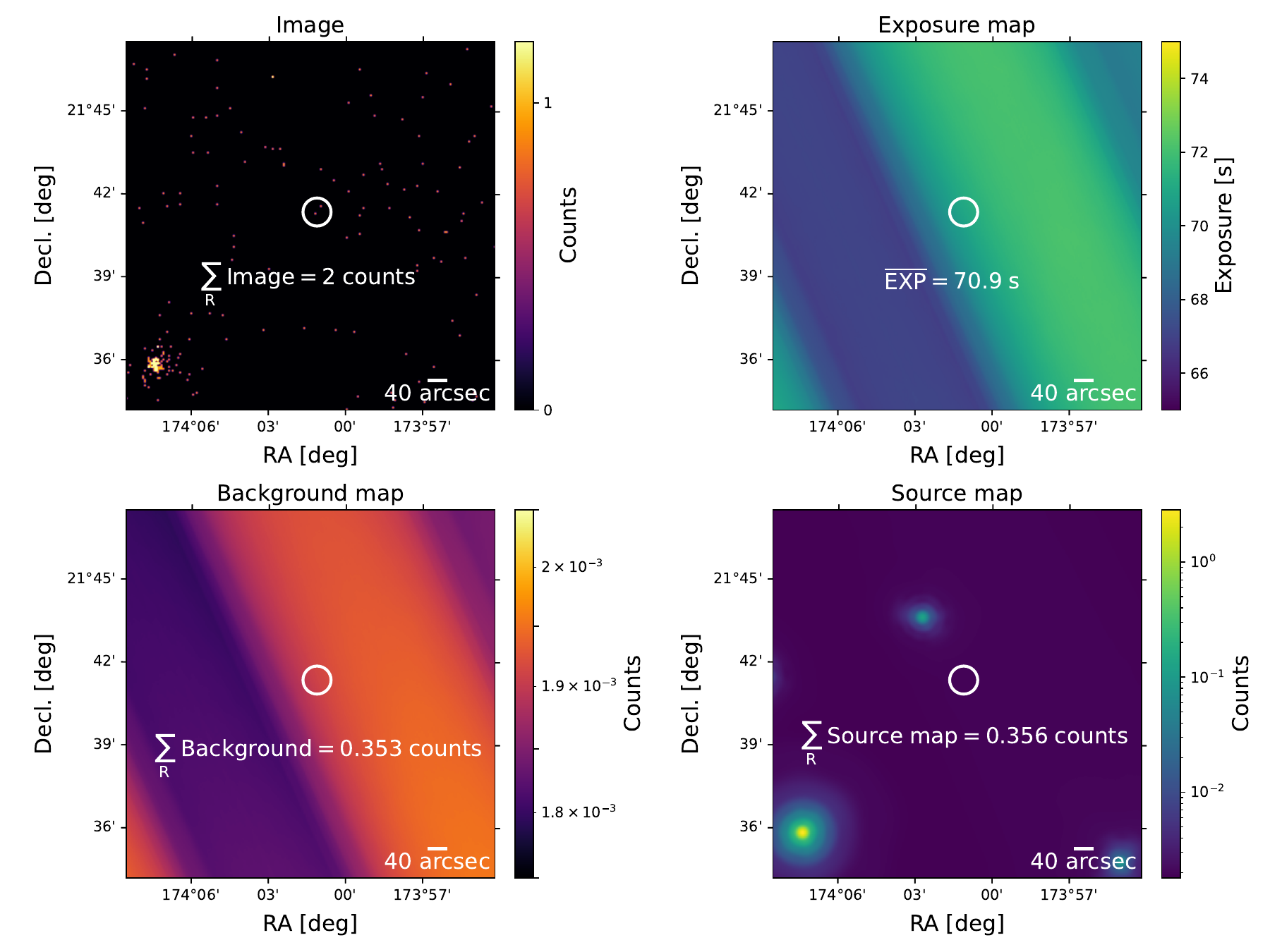}}
      \caption{eRASS1 eROSITA data products for the sky tile 174069 in the $0.2-2.3$ keV energy band. \textit{Top panels:} Science image (\textit{left}) with the discrete observed counts and the exposure time (\textit{right}) image shows the time, in seconds, that each position in the sky was observed on-axis by eROSITA. \textit{Bottom panels:}  Source-free background map (\textit{left}) in units of counts per pixel. The source map (\textit{right}) is produced by the addition of the best PSF-fit model of the detected sources and the source-free background to create a source+background map. In all images, the white circles with a radius of $\sim 30\arcsec$ represent the size of the aperture obtained from an EEF of 75\% of the PSF. The scale bars, shown in the lower right corners of the maps, have an angular size of $40\arcsec$. We note that the background, exposure time, and source map exhibit diagonal stripes produced by the scanning mode of eROSITA. The color scale of the background and exposure images have been selected to highlight this feature. We emphasize that the difference between the higher and lower values of the exposure time and the background maps lie within $\pm4$\% of the mean value.}
         \label{pic:data}
   \end{figure*}

\begin{figure}
   \resizebox{\columnwidth}{!}
            {\includegraphics{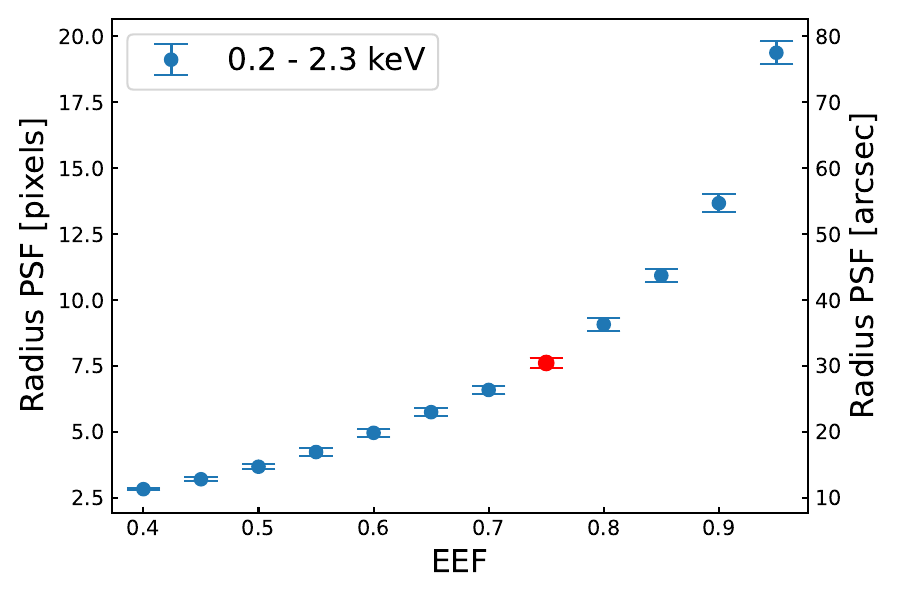}}
      \caption{Visual representation of the shapelet-modeled eROSITA PSF. We show the radius of the survey-averaged shapelet-modeled eROSITA PSF in units of pixels and arcsec as a function of the EEF for the single band detection ($0.2-2.3$ keV). The files with the radius of the PSF are produced by the eSASS task \texttt{apetool} for each eROSITA sky tile. The uncertainties describe the variation of the PSF size within a sky tile. We highlight in red the radius of the PSF at an $EEF=0.75$ that we use for our upper limit calculations which correspond with a radius of $\sim 30\arcsec$.  
      The pixel size of the eROSITA data is 4\arcsec. }
         \label{pic:psf}
   \end{figure}

The \eros data analysis pipeline is organized into task chains related to event calibration, image and exposure map creation, source detection, and the creation of source-specific products such as spectra and light curves \citep[see ][]{2021A&A...647A...1P,2022A&A...661A...1B}. The most relevant standard pipeline data products for the computation of the \eros upper limits are the scientific images and the exposure time maps (see Fig.~\ref{pic:data}). After the event calibration chain \citep[see appendix A.1 of][]{2022A&A...661A...1B}, the eSASS task \texttt{evtool} extracts the detected X-ray photons from the event tables to create images with a pixel size of 4\arcsec. The resulting image is a FITS file that contains the observed counts ($N$). The task \texttt{expmap} computes the exposure time ($t$) defined as the time, in seconds, each position of the sky was covered by the eROSITA field of view. 
 This exposure time is corrected by the vignetting function which depends on off-axis angle and energy. The vignetting-corrected exposure time is stored in the exposure map.

The eSASS task \texttt{erbackmap} is used to compute the background level ($B$) in an iterative process. It first masks out sources from the count image and then applies a two-dimensional adaptive smoothing algorithm to the source-free image during the source-detection chain. The source-free image is produced following the sliding box detection algorithm \texttt{erbox} in local mode with a box size of 7 pixels. From the input image, \texttt{erbox} detects peaks of counts that fall inside the sliding box, computes background counts from the surrounding regions of the box, and calculates the statistical significance of those peaks by adopting a log-likelihood threshold $L=-ln (P_{\Gamma}(N,B))$, where $P_{\Gamma}(a,x)$ is the regularized incomplete gamma function\footnote{$ P_{\Gamma}(a,x) = \int_{0}^{x} e^{-t}t^{a-1}dt / \int_{0}^{\infty} e^{-t}t^{a-1}dt$} \citep[see appendix A.5 of][]{2022A&A...661A...1B}. The significant detections above the threshold are stored in an initial catalog. Their positions are masked out from the count image and a preliminary background map is created with the eSASS task \texttt{erbackmap}. The size of the masked-out sources depends on the count levels of the detected peaks, the PSF, and the source extension. Since a very low detecting threshold is used in the standard eSASS pipeline, background fluctuations might be detected as spurious sources and masked from the image. Thus, the smoothed background might be biased toward low count levels. A new sliding-box detection is run, now considering the preliminary background map and the source positions of the initial catalog. This new iteration updates the log-likelihood of the detections, reduces the number of spurious sources, and creates an updated catalog with a less biased background. Another final \texttt{erbox} iteration creates a new updated catalog and background image which are then passed on to the PSF-fitting algorithm \texttt{ermldet} to determine the final (cataloged) source parameters. The final background map is shown on the bottom-left panel of Fig. \ref{pic:data}.

Since \texttt{erbackmap} masks sources with a log-likelihood above a certain threshold to perform the adaptive smoothing of the source-free image, we test the dependence of the background level on this log-likelihood threshold. 
The eSASS pipeline (c010 version) uses a log-likelihood threshold of $\rm ML=6$ to identify sources when creating the background. We run \texttt{erbackmap} considering log-likelihoods of 5, 10, 15, and 20. We find that the $\rm ML=5$ background maps have an average count level 4\% lower than that based on an $\rm ML=6$ run. For $\rm ML=10$, 15, and 20, the background is on average 8\%, 11\%, and 13\% higher than for the $\rm ML=6$ run, respectively.
Interestingly, when we compare the results based on the background from $\rm ML=6$ and $\rm ML=20$, this translates to an upper flux-limit discrepancy that is less than 1\%. This extremely low dependence on the background level (thus the limiting log-likelihood of masked detections) is reasonable because the background level is low compared to the observed counts ($B<<N$). Thus, a change of the background level even by $\sim$10\% does not change the ratio between observed counts and background counts significantly. 

The PSF-fitting algorithm \texttt{ermldet}, used to determine the final source properties, can also provide valuable information for the upper limit server. \texttt{ermldet} applies a maximum likelihood PSF-fitting procedure to all the sources of the input list produced by the sliding box algorithm \citep[see appendix A.5 of][for more details]{2022A&A...661A...1B}. 
In addition to the best-fit position, extension, counts, and count rates of the detected sources, it also produces source maps. The source map consists of the best-fit PSF models of all detected sources added to the smoothed background (bottom-right image of Fig. \ref{pic:data}). The source maps provide an alternative data set to be considered as background when calculating upper limits in regions close to a bright and detected source. In such regions, the user might be interested in the upper limit calculation of a source that lies on top of a detected source, such as the study of diffuse emission contaminated by foreground stars. Thus, by using the source map, the counts from the foreground star will be considered background counts, obtaining an upper flux limit that indicates the maximum flux that the diffuse emission can have. We note that this is in addition to the option of using the source-free background maps (produced by \texttt{erbackmap}).

\subsection{Choosing the EEF}\label{choosingeef}

To use the eROSITA data products to calculate our upper limits, we performed aperture photometry. The aperture photometry on the image, background image, and exposure map is performed by considering a circular aperture with a radius $R$. This radius is calculated based on the size that encloses a given encircled energy fraction (EEF) of the PSF in a given energy band. The EEF can be expressed with the following equation:

\begin{equation}
    \rm EEF = \left. \int_{0}^{R} PSF(E,r,\theta)dr \middle/ \int_{0}^{\infty} PSF(E,r,\theta)dr, \right.
\end{equation}

\noindent where the PSF is integrated radially until the radius, $R$, such that the desired fraction of the total PSF ($\rm EEF=0.75$) is reached.

The eROSITA PSF has an approximately regular Gaussian profile near the optical axis and elongated and asymmetric features at larger off-axis angles. As the eROSITA all-sky surveys are made in scanning mode, the PSF for each sky position will be an average of on-axis and various off-axis angles, as the position passes through eROSITA's field of view. This average PSF is not as small as if a source had only been observed at the on-axis angle, but it is approximately constant during the all-sky scans. Thus, the PSF shape does not depend on the sky position anymore but only on the energy band. 

The eROSITA PSF can be represented as images or shapelet models \citep[see Appendix B.1 of][for more details]{2022A&A...661A...1B}. These PSFs were experimentally produced in the ground calibration at the MPE PANTER\footnote{\url{https://www.mpe.mpg.de/heg/panter}} facility at different energies using an X-ray point source located 124 m away from the mirror assembly. The shapelet functions are a two-dimensional set of orthonormal weighted Hermite polynomial functions that correspond to perturbations of a circular Gaussian profile \citep{2003MNRAS.338...35R}. The shapelet coefficients are determined by fitting the PANTER PSF images with the shapelet models at different energies and detector positions.

At a given energy range, the eSASS task \texttt{apetool} combines the shapelet coefficients with the attitude file (event file and good time intervals) of a given observation to generate a model of the exposure-time averaged PSF shape and stores them in calibration files. These files correspond with grids that describe the variations of the PSF size, in units of pixels, across the sky tile field of view. This grid is produced for EEFs in the range 40--95\% in steps of 5\%. The results are stored in a three-dimensional data cube. The density of the grid is a trade-off between speed, size of the final PSF data product, and an adequate description of the variations of the PSF size across the field of view \citep[][]{2022A&A...661A...1B}. The default setup is a grid of $21\times21$ positions along the X and Y axis, resulting in a final $21\times21\times 12$ data cube containing the radius of the PSF as a function of the position on the sky and EEF. Using these data, the radius of any desired aperture can be computed directly from an interpolation of this three-dimensional (3D) map. Figure \ref{pic:psf} shows the average size of the PSF obtained from the shapelet modeling of the eROSITA PSF, as a function of different EEFs. The data points and uncertainties in the image correspond to the average and standard deviation of the $21\times21$ PSF radius at each EEF, respectively. We note that the radius of the PSF increases at larger EEF because of the elongated wings.

In order to select the optimal EEF and thereby the size of the aperture to be used for X-ray aperture measurements, we considered 100,000 random positions in the sky tile 174069. From those positions, we collected observed counts using different EEF ranging between 0.6 to 0.95 in steps of 0.05. We divided the observed counts by the fraction of the PSF used in order to correct for the photons that fall outside the aperture. We note that the corrected observed counts do not vary significantly between an EEF of 0.6 and 0.8, suggesting that the number of observed counts does not depend on the specific choice of the EEF within this range. Since the eSASS pipeline also performs aperture photometry for the detected sources (independent from the PSF fitting), we chose a value of EEF=0.75 to be consistent with the main pipeline. For the selected energy band $0.2-2.3$ keV, the aperture has a size of $R = 7.6 \pm 0.2 $ pixels ($30.5 \pm 0.8$ arcsec), which is highlighted with the red data point in Fig. \ref{pic:psf}. 

To test the impact of the uncertainty on our aperture size, we performed aperture photometry and calculated the upper limits using the 1$\sigma$ upper and lower boundaries on the nominal radius for EEF=0.75. Specifically, we used $R_{\rm lower}= 7.4$ pixels and $R_{\rm upper}=7.8$ pixels. The number of observed counts shows a difference of less than 1\% when using $R_{\rm lower}$ or $R_{\rm upper}$. We conclude that the precise choice of $R$ does not have a significant impact on the upper limit calculation and we therefore used a radius defined by 75\% of the PSF.

\subsection{X-ray aperture photometry}\label{photometry}
The aperture photometry and the subsequent upper limit computations are coded in \textsc{python} \citep{10.5555/1593511}. The analysis consists of collecting the observed counts, background counts, source-map counts, and average exposure time at every pixel of the eROSITA-DE footprint using an aperture with a radius $R$, equivalent to an EEF of 0.75. 
Since the aperture photometry is performed in pixel space, some pixels might not be fully covered by the circular aperture. We test two methods to measure the number of counts from pixels within the circular aperture. 
In the first case, we consider the central position of the pixel. If the center of a given pixel falls inside the circular aperture, the data of the whole pixel will be considered for the upper limit calculation (referred to as the ``center'' method). 
The second method considers fractional pixels. If the pixel is partially covered by the aperture, only a fraction of that pixel will be used to calculate the upper limits. 
This fraction is computed based on the area of the pixel that is covered by the aperture, regardless of whether the center is included or not. We find that, on average, both methods collect the same number of counts and background counts, and the mean discrepancy in the final upper limit value is lower than 1\% between the two methods. We, therefore, decided to use the ``center'' method to do the aperture photometry. Once we collect the counts within the aperture, the photometric data are tabulated for the position of interest. 
We retrieve all the photometric products at every eROSITA pixel by convolving\footnote{Convolution routine from Python package \textsc{scipy.signal.oaconvolve}} the initial data products (observed counts, background, and exposure maps) with the circular aperture. The convolution process speeds up the aperture photometry considerably compared to a pixel-by-pixel photometry routine. We note that the eSASS task \texttt{apetool}, part of the eROSITA detection chain, also computes aperture photometry products (counts, background counts, mean exposure time) in addition to the PSF maps described in \S \ref{choosingeef}, and sensitivity maps. We compared our Python routine with \texttt{apetool} by collecting the photometric data at the positions of the detected sources of the first eROSITA catalog \citep{Merloni2024}. We note that the observed counts obtained with both methods follow a one-to-one relationship with a small scatter and a standard deviation of $\sim 7\%$. We also note that a discrepancy of $\sim 7\%$ in the counts only leads to a discrepancy lower than $\sim4\%$ in the final upper limit. Therefore, the main reason for choosing the Python approach is the speed and flexibility of collecting photometry products at every pixel and sky tile of the eROSITA footprint. 
Finally, once the data are collected, the upper flux upper limits are calculated using Eq. \ref{eq:fl}, where the ECF is calculated based on an absorbed power law model with a photon index of $\Gamma=2.0$ and a column density of $N_{\rm H}=3\times10^{20}$ cm$^{-2}$ (see \S \ref{sec:ecf}).

\subsection{Details on the upper limit products}\label{products}

Once we collected all the photometric data from a sky tile, we store them efficiently in order to optimize future positional searches. Following the methodology of RapidXMM \citep{2022MNRAS.511.4265R}, we used the Hierarchical Equal Area Iso Latitude pixelation\footnote{\url{http://healpix.jpl.nasa.gov/}} of the sphere \citep[HEALPix;][]{2005ApJ...622..759G} to create a unique index per pixel. It takes into account the position of the pixel projected on the sky and creates a one-to-one relation between the input coordinate (R.A. and Dec.) and the HEALPix index. The R.A. and Dec. are obtained from the world coordinate system (WCS) of each sky tile. To convert from coordinates to HEALPix index, we use the task \texttt{skycoord\_to\_healpix()}, which is part of \texttt{astropy-healpix}\footnote{\url{https://astropy-healpix.readthedocs.io/}}, the Python implementation of the HEALPix algorithm. In order to have one unique HEALPix index associated with each eROSITA pixel, we consider a HEALPix tesselation of order 16 (\textsc{NSIDE}=$2^{16}$), which creates HEALPix cells with a resolution of $\sim$3\arcsec, similar to the eROSITA pixel size of 4\arcsec. Each HEALPix index contained within a sky tile is then associated with an upper flux limit and the corresponding photometric products. 

In addition to the calculated upper limits, we store the aperture-based total counts, background counts, and exposure time for every pixel. This is with the aim of providing as much flexibility as possible for future users. With these data, Eq.~\ref{ul} can be used to compute upper limits at any confidence level other than the pre-computed (one-sided 3$\sigma$). With the aperture-based observed counts and background counts, it is also possible to reproduce the posterior distribution function of Eq.~\ref{poisson} and estimate source fluxes with asymmetric errors at any confidence interval by following the methodology of \S \ref{sec:theory} and \cite{1991ApJ...374..344K}.

\begin{figure}
\centering
\includegraphics[width=\columnwidth]{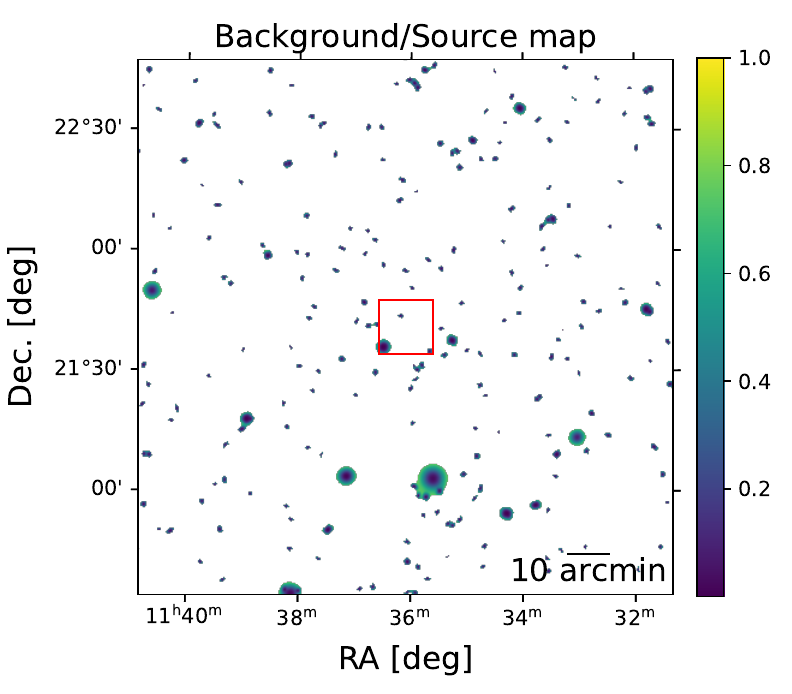}
\caption{Ratio between the background map (source-free) and the source map (detected sources + background). Pixels, where the ratio is smaller than 0.8, are displayed. This map is used to assign the positional flag and indicate that the upper limit calculation will be affected by detected sources. 
If the area used for the aperture photometry intersects or includes such a region, the ``close neighbor'' flag is set to 1 for this pixel, otherwise 0. The red square highlights the sky region displayed in Fig.~\ref{pic:data}. }\label{pic:sources}
\end{figure}

To make the user aware that a particular sky position is close to a bright source, we take advantage of the source map to create positional a flag that indicates whether the position of interest is close to a detected source or is in a source-free region. This flag is called ``close neighbor'' and it is obtained as follows: 
We define a threshold of $0.8$ on the ratio between the background counts and the source map counts. An example of this ratio, for the sky tile 174069 is shown in Fig.~\ref{pic:sources}. If this ratio is larger than 0.8, the background and source map values are rather similar, implying that the position of interest is not close to a detected source. We note that when the background is at the level of $80\%$ or higher of the counts from the source map, the corresponding flux upper limits have a discrepancy of less than $\sim$1\% between the upper limit based on the background map (source-free) and the source map (background + detected sources).  
Any pixel with a ratio lower than the threshold, namely, between 0.0 and 0.8, 
indicates that the aperture photometry of this pixel overlaps with a detected source (or is very close to it, so that the source influences the aperture measurements by $>$1\%). All these pixel positions will be given the ``close neighbor'' flag 1. 
We used the map (see Fig.~\ref{pic:sources}) as a flag decision mask that indicates the regions in which the aperture will be affected by the presence of neighboring (detected) sources.
We emphasize that the close neighbor flag is a warning to indicate cases when the area used for the aperture photometry contains a detected source(s). 
The details of the calculation of the upper limits and photometric products do not change when the flag is activated and we do not subtract the counts of detected sources (as given by the source map).

As mentioned in \S \ref{erodata}, we also computed the aperture photometry based on the source map image. The source map consists of a map with the best model of the detected sources added to the smoothed background in units of counts per pixel. The source map counts can be considered as an alternative ``background'' for the upper limit calculation when the user is interested in the upper limit of a hypothetical second source that lies on top or close to a detected source. Thus, by using the source map, the upper flux limit will indicate the maximum flux that the second source could have. This is in addition to the option of using source-free background maps. As shown in the example aperture included in Fig.~\ref{pic:data}, the source map counts within the aperture are slightly higher than the background counts. Although the aperture is located in a source-free region, the outer wings of the PSF model of a detected source still account for an increase in the summed counts. In the example shown in Fig.~\ref{pic:data}, the difference between the upper limit using the source map and the (source-free) background map is $\ll 1$\%. Following our flagging algorithm, the central pixel of interest will not be flagged (``close neighbor'' flag = 0) as it is still considered to be a source-free pixel.

\begin{table*}
\small\addtolength{\tabcolsep}{+1.0pt}
\caption{Description of the output tables per sky tile.} 
\label{table:1} 
\centering
  
\begin{tabular}{llll}   
\hline
\hline
Column name & Format\tablefootmark{a} & Units & Description \\   
\hline                                   
HEALPix index & K & -- & Projection of the coordinates into HEALPix map. \\ 
Counts & J & cts &Extracted counts within the aperture from the science image.\tablefootmark{a}  \\
Bkg\_counts & E & cts & Extracted background counts within the aperture from the background image.\tablefootmark{b}  \\
Bkg\_SourceMap & E & cts &Extracted counts within the aperture from the source-map image.\tablefootmark{b}  \\
Exposure & E & s & Average exposure time within the circular aperture. \\
Flag\_pos & I & -- & Close neighbor flag\tablefootmark{c}.  \\
Flux\_UL\_B & E & $\rm erg\; s^{-1}\; cm^{-2}$ & Flux upper limit at CL=0.9987 (one-sided) based on column ``Bkg\_counts''.\tablefootmark{d} \\
Flux\_UL\_S & E & $\rm erg\; s^{-1}\; cm^{-2}$ & Same as UL\_B, but based on the source-map counts (``Bkg\_SourceMap'').\tablefootmark{d} \\
\hline
\end{tabular}
\tablefoot{
~\tablefootmark{a}{Format of the columns as defined in \url{https://docs.astropy.org/en/stable/io/fits/usage/table.html#column-creation}. In particular, E denotes a single precision float (32-bit), I a 16-bit integer, J a 32-bit integer, and K a 64-bit integer.}
~\tablefootmark{b}{Extracted from an aperture defined by an EEF=0.75 of the PSF.}
~\tablefootmark{c}{Set to 1 if the aperture of a given pixel overlaps with a detected source and 0 in source-free regions, according to the procedure presented in \S \ref{products}.}
~\tablefoottext{d}{Corrected by the fraction of the PSF (see Eq.~\ref{eq:fl}). Flux obtained for an absorbed power-law spectral model with a photon index of $\Gamma=2.0$ and a column density of $N_{\rm H}=3\times10^{20}$ cm$^{-2}$. The header of the FITS file contains the ECF of the absorbed power law model. Thus, the user can multiply the upper flux limit with this factor to recover the upper limit in units of counts per second and apply any other ECF for their preferred spectral model (see \S \ref{sec:ecf}). }
}
\label{maintable}
\end{table*}

\begin{figure}
\includegraphics[width=\columnwidth]{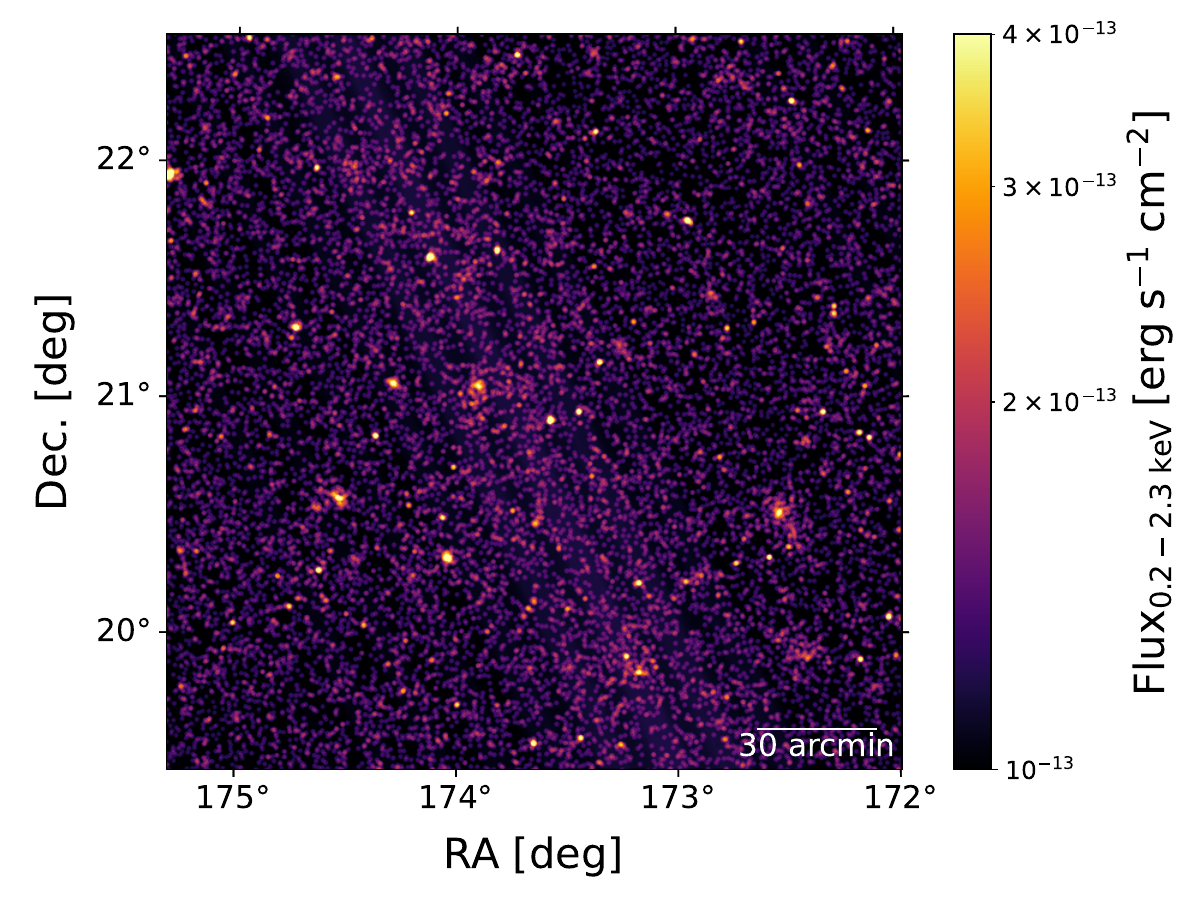}
      \caption{Reconstructed spatially resolved upper flux limit image of the sky tile 174069 for the single band $0.2-2.3$ keV using a confidence interval of 99.87\% (one-sided). For the computation, we used the source-free background map.} 
         \label{pic:ulmap}
   \end{figure}

Table~\ref{table:1} summarizes the columns of the final eROSITA upper limit products. For each pixel in a given sky tile at a certain energy range, our routine stores a table entry with the following columns: The HEALPix index, the integrated observed counts, background counts, source-map counts, mean exposure time inside the aperture, the close neighbor flag, and the upper flux limits at the one-sided $3\sigma$, corresponding to a confidence level of $\rm CL=0.9987$ in a normal distribution as used in Eq. \ref{ul}. The reported upper flux limits are already corrected by the aperture fraction (i.e., $\rm UL_{reported} = UL_{eq. \ref{ul}}/EEF_{0.75}$) and they were calculated using an absorbed power-law spectral model with a photon index of $\Gamma=2.0$ and a column density of $N_{\rm H}=3\times10^{20}$ cm$^{-2}$ to be consistent with the spectral model used in the main source catalog. We note that the upper flux limits come in two flavors: the first is based on the source-free background, while the second uses the source maps (background + sources). Both upper limits will be provided. As an example of the upper flux limit, Figure~\ref{pic:ulmap} shows the spatially resolved upper flux limit map with a confidence interval of 99.87\% (one-sided) for the sky tile 174069 based on the background map. The typical upper flux limits in source-free regions reach values down to $\rm 1\times 10^{-13}\; erg\; s^{-1}\; cm^{-2}$. 

\begin{figure*}
\includegraphics[width=\textwidth]{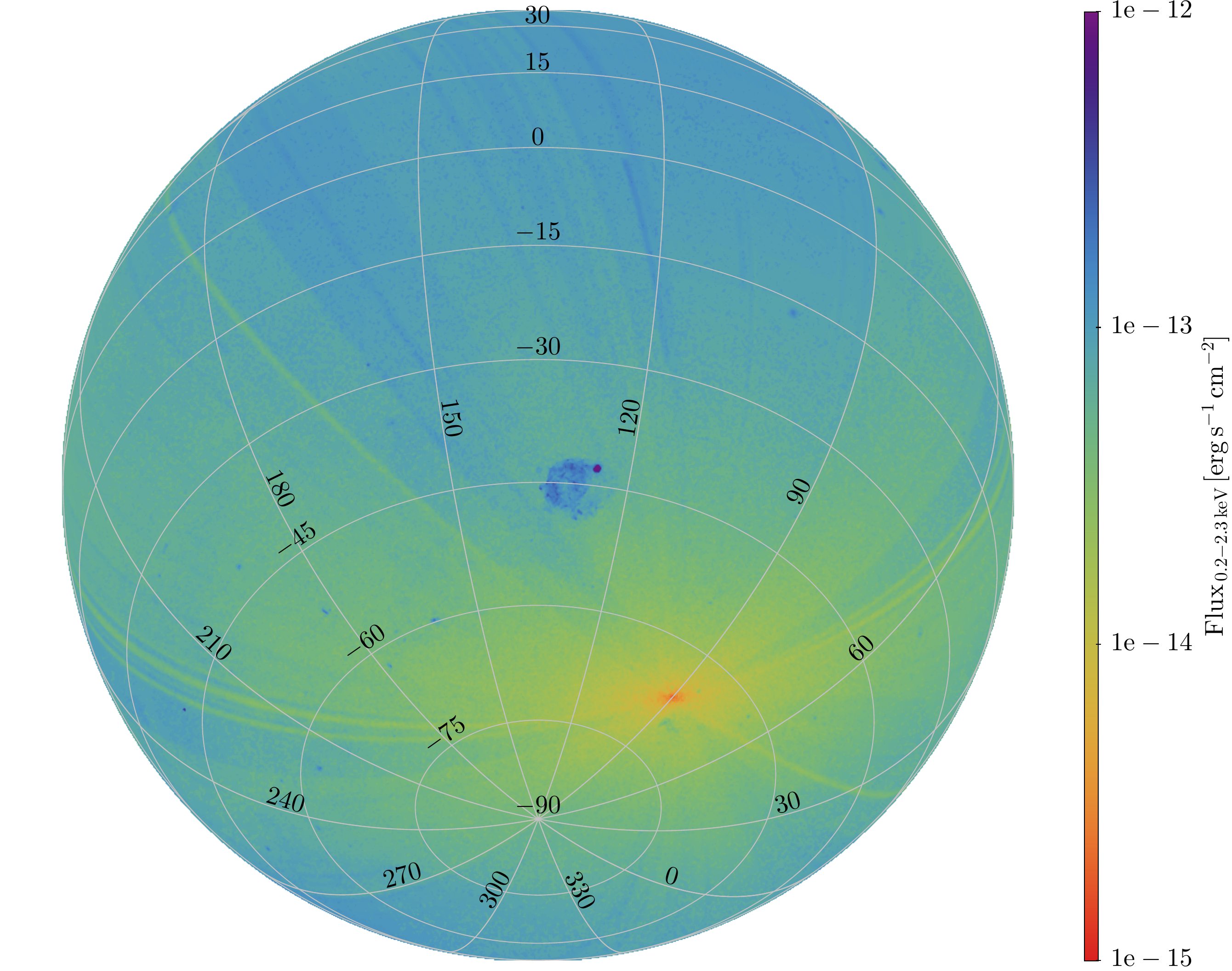}
      \caption{All-sky upper limit map. We show the flux upper-limit map for eRASS1 (German sky) for the single-band detection run in the $0.2-2.3$ keV energy band. The upper limits are computed using a confidence interval of 99.87\% and a spectral model consisting of a photon index of $\Gamma=2.0$ affected by galactic absorption of $N_{\rm H}=3\times10^{20}$ cm$^{-2}$. The map is plotted in orthographic projection where the meridian gray lines correspond to R.A. and the parallel lines to Dec. }
         \label{pic:ALLmap}
   \end{figure*}

We present an $0.2-2.3$ keV upper flux limit map (based on the source-free background counts) for the entire German eROSITA sky in Fig.~\ref{pic:ALLmap}. The lowest upper flux limits ($\rm \sim 1\times 10^{-15}\; erg\; s^{-1}\; cm^{-2}$) are found around the Southern ecliptic pole which receives the deepest exposure due to the eROSITA scanning strategy. In the equatorial plane, the upper flux limit is a few times $\rm10^{-13}\; erg\; s^{-1}\; cm^{-2}$.

\subsection{Upper limits at other energy bands}

In addition to the single band source detection, eSASS pipeline also runs a source detection simultaneously in three energy bands that cover the soft ($0.2-0.6$ keV), medium ($0.6-2.3$ keV), and hard ($2.3-5.0$ keV) bands. The three-band source detection follows the same steps mentioned in Sect. \ref{erodata}, and it is optimized to detect sources with soft or hard spectra. In particular, if a soft source is next to a hard source, the three-band source detection run has a higher chance of being able to resolve both sources than a single-band run.

\begin{figure}
    \centering
    \includegraphics[width=\columnwidth]{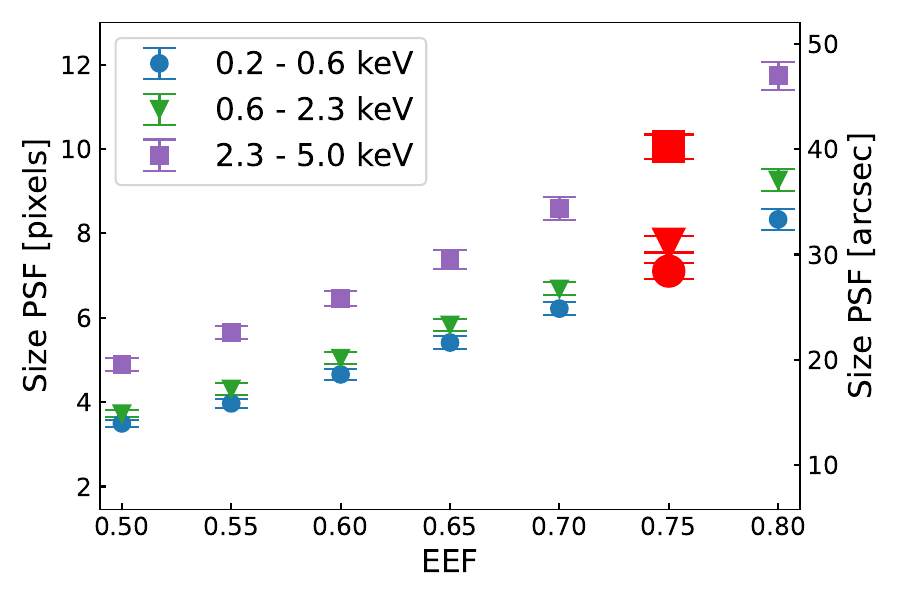}
    \caption{Size of the shapelet-modeled eROSITA PSF in units of pixels and arcsec as a function of the EEF in the soft (blue circles: $0.2 - 0.6$ keV), medium (green triangles: $0.6 - 2.3$ keV), and hard (purple squares: $2.3 - 5.0$ keV) energy bands. Similar to Fig. \ref{pic:psf} but based on the three-band detection pipeline. The data and error bars, like in Fig. \ref{pic:psf}, correspond to the mean and standard deviation of the 2D $21 \times 21$ array that models the size of the PSF using shapelet coefficients as a function of EEF. We highlight with red and bigger markers the size of the PSF at an EEF$=0.75$ used for the aperture photometry. For the purpose of visualization, we only plot data points between EEF 0.5 and 0.8.}
    \label{pic:sizeeef_3b}
\end{figure}

Figure~\ref{pic:sizeeef_3b} shows the size of the PSF for these three energy bands as a function of the EEF. We note that the size of the PSF increases moderately as a function of energy. As for the single detection run in the $0.2-2.3$ keV energy band, we use an $\rm EEF=0.75$ that corresponds to a size of $7.1 \pm 0.2$, $7.7 \pm 0.2$, and $10.1\pm 0.3$ pixels for the soft, medium, and hard band, respectively.

Since the German eROSITA team does not run a single source detection in the $0.2-5.0$ keV band, the pipeline does not provide individual data products in this band on which we can perform aperture photometry. Thus, we assume that the aperture-collected counts of the individual sub-bands can be added to obtain the total observed counts and background counts of the $0.2-5.0$ keV band:
\begin{equation}
\begin{split}
N_{0.2 - 5.0\; \rm keV}&= N_{0.2 - 0.6\; \rm keV}+N_{0.6 - 2.3\; \rm keV}+N_{2.3 - 5.0\; \rm keV}, \\
B_{0.2 - 5.0\; \rm keV}&=B_{0.2 - 0.6\; \rm keV}+B_{0.6 - 2.3\; \rm keV}+B_{2.3 - 5.0\; \rm keV}.
\end{split}
\end{equation}

Despite being collected with different aperture radii, as shown in Fig.~\ref{pic:sizeeef_3b}, the considered fraction of the PSF is always the same, which means that the collected counts always correspond to the counts that fall within 75\% of the PSF in the sub-bands. Finally, the upper limit of the summed three-band, in units of counts, is given by: $UL_{0.2 - 5.0\; \rm keV} = UL(N_{0.2 - 5.0\; \rm keV},B_{0.2 - 5.0\; \rm keV})/0.75$. We use the exposure time of the most sensitive energy band ($0.6-2.3$ keV) as the exposure of the summed three-band in order to compute count rates and X-ray fluxes. 

\section{Upper flux limit with different spectral models}\label{sec:ecf}

In this section, we discuss the effects of choosing the correct spectral model for the final flux upper limits computation. This is relevant if the sources of interest are known to have very different spectra from the standard power law. These sources could include extremely soft spectra as in neutron stars, thermal spectra of stars, or tidal disruption events. Our final upper-limit data products contain the tabulated aperture counts and exposure time (see Table~\ref{maintable}) which can be used as an input for Eq.~\ref{ul} to derive the upper limits in units of counts at any particular confidence interval. Using Eq.~\ref{eq:fl}, one can correct for the desired input spectrum, which is controlled by the ECF value. A detailed and simple description of how to calculate various ECF is presented in Appendix \ref{a1}, while Tables \ref{ecftable} and \ref{ecftable3B} present several ECF based on different spectral models. The tables also show the parameters used for the models, the ECF, and a multiplication factor that corrects the tabulated upper flux limit to match the chosen spectral model.

We note that in the $0.2-2.3$ keV band, there is a maximum discrepancy\footnote{$\rm (ECF_{pow,2.0} - ECF_{bb,0.05})/ECF_{pow,2.0}$} of approximately 29\% between the upper flux limits when we consider the black-body model with $kT=0.05$ keV and a power law with $\Gamma=2.0$. This discrepancy is caused by the choice of the ECF. Consequently, the upper flux limits based on a power-law model are 29\% smaller than when using a black-body emission. We conclude that the ECF introduces the strongest uncertainty in the flux upper limit calculation. Any other decision, such as the size of the aperture, the log-likelihood of the background level, or the method of collecting counts in pixel space, is negligible compared to the impact of the spectral model. This emphasizes the importance of the right choice of the spectral model and the need of recomputing the upper flux limits from the tabulated aperture counts and exposure time if the source model is different from a power-law model with $\Gamma=2.0$. If a different confidence level is of interest to the user, such upper flux limits can be also calculated from Eq.~\ref{ul} and the input values, as given in Table~\ref{maintable}.

\section{Access to the upper limit data}\label{sec:webtool}

The upper flux limit data can be accessed in two ways. One can either download the data table formatted as shown in Table~\ref{table:1} or access the data through a web tool. We recall that the precomputed upper flux limits (column UL\_B and UL\_S) refer to a one-sided confidence interval of CL = 99.87\% (corresponding to a one-sided 3$\sigma$ interval) and assume a spectral model with $\Gamma=2.0$ and $N_{\rm H}=3\times10^{20}$ cm$^{-2}$. If a different confidence interval or a different spectral model is preferred, the upper flux limits should be recomputed by following Eqs. 5 and 7 and using the observed counts, background (or source map), and exposure time columns of the table. To enable this analysis for all users, both access methods provide the required additional data to make these calculations. In synchronization with the official eROSITA data release (DR1), only data based on the eROSITA-DE sky are available for eRASS1. 

\subsection {Downloading the data}
The data\footnote{\url{https://erosita.mpe.mpg.de/dr1/erodat/data/download/}} are stored in several identically formatted tables. Each contains the flux limits at different energy ranges. The name of our upper limit tables follows the same name scheme as the eROSITA products\footnote{The eROSITA naming convention is described in \url{https://erosita.mpe.mpg.de/dr1/eSASS4DR1/eSASS4DR1_ProductsDescription/file_naming_scheme_dr1.html}}.

For future planned data releases, eROSITA-DE will make further eROSITA all-sky scans available, including upper flux limits based on stacking data from several eROSITA surveys and matching the updated data processing. In order to access the upper-limit information for a certain position on the sky, the user will be required to conduct two calculations: i) determine in which eROSITA sky tile the input position is located and ii) obtain the HEALPix index in the specific sky tile that corresponds to the required R.A. and Dec. Finally, the corresponding upper limit can be retrieved by identifying the row of the computed HEALPix index (for the sky position of interest) in the downloaded table of the sky tile. A technical description of the nomenclature of the eROSITA upper limits files and how to retrieve the upper flux limit for a particular set of coordinates within the downloaded table are provided in Appendix \ref{appendix:retrievingul}. 

\subsection {The eROSITA upper limit server}
The data given in Table~\ref{table:1} can also be accessed via a web tool\footnote{\url{https://erosita.mpe.mpg.de/dr1/AllSkySurveyData_dr1/UpperLimitServer_dr1/}}. This eROSITA upper limit server will not only provide the upper flux limits for a sky position (in R.A. and Dec.; limited to the German eROSITA sky) but also gives the eRASS catalog entry for the nearest detected sources when the close-neighbor flag is triggered for the sky position of interest. 
Providing not only the upper flux limits but also the detection properties of the closest neighboring sources is advantageous when the user is unaware of the detected sources in the field of interest.
In the case that a detected source is close by, there can be a significant difference between upper flux limits from the columns UL\_B and UL\_S. We note that we ought to keep in mind the different scientific interpretations of these two upper flux limit values (see \S \ref{products}).

Details on the upper limit server and example queries will be provided on the webpage. The layout of our database follows the design of the RapidXMM upper limit server \citep{2022MNRAS.511.4265R}, where the input sky coordinates are transformed to HEALPix indexes and then matched with the indexes stored in the tables.

\section{Conclusions}\label{sec:conclusions}
We present the upper flux limits for the first all-sky eROSITA scan (limited to the German sky half: $180^{\circ}\lesssim l \lesssim 360^{\circ}$). The limits are derived by using X-ray aperture photometric measurements and the Bayesian approach described by \cite{1991ApJ...374..344K}. 

The upper limits are computed for every pixel position in the eROSITA scan at a confidence interval of CL = 99.87\% (corresponds to a one-sided 3$\sigma$ level). Two different options for upper limits are available: i) upper limits that are computed using the background map (source free; UL\_B) and ii) upper limits that use the source map (background plus best-fit models of detected sources; UL\_S) as a background measurement.
These upper limits should be interpreted and used as follows:
\begin{itemize}
    \item{In source-free regions (close neighbor flag = 0), UL\_B represents the maximum flux of a non-detected source at the given position, based on the observed counts and the background level. We note that in source-free regions, UL\_B and UL\_S are virtually identical.} 
    \item{If the close neighbor flag for a requested position is set to 1, UL\_S (background plus best-fit PSF models of detected sources) should be used. Since detected sources in the proximity are considered as background, this upper flux limit should be interpreted as the maximum flux of a hypothetical second source in addition to the already detected source. The values from UL\_B should not be used in these cases, and UL\_B should not be interpreted as the upper limit for the hypothetical case where the detected source was not there.} 
    \item{If the position of interest coincides with a detected source (or its PSF), the use of UL\_S and its interpretation for a hypothetical second source is still valid. For rather bright sources, UL\_B can be interpreted as the one-sided 3$\sigma$ upper flux uncertainty of the source. Indeed, our tabulated input values for the upper-limit calculation can be used to recover the posterior distribution of Eq. \ref{post} and thereby numerically compute the asymmetric lower and upper flux uncertainties of the sources, as described in \cite{1991ApJ...374..344K}. We note that neither UL\_B nor UL\_S can be used to evaluate how sensitive eROSITA would have been at the position of interest if no source were there. The upper-limit values at the position of a detected source will always be larger than the flux of the source itself.} 
\end{itemize}

We computed upper limits for a single-band source detection using an energy band of $0.2-2.3$ keV, as well as in all sub-bands of a three-band source detection run (soft: $0.2-0.6$ keV, medium: $0.6-2.3$ keV, hard: $2.3-5.0$ keV). We also combined all three sub-bands and we give upper limits for the $0.2-5.0$ keV band. All upper-limit data are stored with hierarchical indices (HEALPix) framework to enable a fast search. We describe how to convert from sky position (RA and Dec.) to hierarchical index and vice versa. 
The data can be accessed by downloading the upper-limit products, one file for each energy range, or by using the eROSITA upper-limit server. 

Throughout the paper, we estimate the potential uncertainties and discrepancies of each decision on how to compute the upper limits. We note that by far the largest uncertainty comes from the assumed spectral model of the source. 
Our precomputed upper flux limits use
a model with $\Gamma=2.0$ and a galactic absorption of $N_{\rm H}=3\times10^{20}$ cm$^{-2}$. Substituting the power-law component of the model by a black-body model with $kT=60$\,eV in the $0.2-2.3$ keV energy band yields about 30\% higher flux limits.  
Therefore we describe and recommend recomputing upper flux limits based on the preferred spectral model. 
To do so, our data products include all necessary input values at each pixel position, namely: the aperture computed observed counts, background counts, and mean exposure time.
We plan to produce similar upper-limit products for future releases of additional eROSITA scans, including upper limits based on stacks of multiple scans.

\begin{acknowledgements}
We thank the referee for their useful comments that helped to improve the manuscript. We thank Takamitsu Miyaji for the interesting discussion on upper limits. D.T. acknowledges support by DLR grant FKZ 50 OR 2203. 
M.K. acknowledges support from DFG grant number KR3338/4-1.
G.L. acknowledges support from the German DLR under contract 50 QR 2104.
D.H. acknowledges support from DLR grant FKZ 50 OR 2003
A.G. acknowledges support from the EU H2020-MSCA-ITN-2019 Project 860744 ``BiD4BESt: Big Data applications for Black hole Evolution Studies'' and the Hellenic Foundation for Research and Innovation (HFRI) project ``4MOVE-U'' grant agreement 2688.
K.P. acknowledges support from the German \textit{Leibniz Community} under program 67/2018.
I.T. acknowledges support from DLR grant 50 OX 1901. 
O.K. acknowledges funding by the Deutsches Zentrum f\"ur Luft- und Raumfahrt contract 50 QR 2103.
This work is based on data from eROSITA, the soft X-ray instrument aboard SRG, a joint Russian-German science mission supported by the Russian Space Agency (Roskosmos), in the interests of the Russian Academy of Sciences represented by its Space Research Institute (IKI), and the Deutsches Zentrum f\"{u}r Luft- und Raumfahrt (DLR). The SRG spacecraft was built by Lavochkin Association (NPOL) and its subcontractors, and is operated by NPOL with support from the Max Planck Institute for Extraterrestrial Physics (MPE). The development and construction of the eROSITA X-ray instrument was led by MPE, with contributions from the Dr. Karl Remeis Observatory Bamberg \& ECAP (FAU Erlangen-Nuernberg), the University of Hamburg Observatory, the Leibniz Institute for Astrophysics Potsdam (AIP), and the Institute for Astronomy and Astrophysics of the University of T\"{u}bingen, with the support of DLR and the Max Planck Society. The Argelander Institute for Astronomy of the University of Bonn and the Ludwig Maximilians Universit\"{a}t Munich also participated in the science preparation for eROSITA. The eROSITA data shown here were processed using the eSASS/NRTA software system developed by the German eROSITA consortium.
\end{acknowledgements}

\begin{appendix}

\section{ECF calculation}\label{a1}

In this section, we present a straightforward recipe to compute ECF and therefore upper flux limits for any preferred spectral model. One way to obtain ECFs is using the X-ray Spectral Fitting Package \citep[XSPEC;][]{1996ASPC..101...17A}. Our base-line spectral model for all upper-limit calculations so far considers a model of an absorbed power-law \texttt{tbabs*pow} with a column density $N_{\rm H}=3\times10^{20}$ cm$^{-2}$ and a photon index of $\Gamma=2.0$.

Obtaining ECFs with XSPEC requires the eROSITA specific redistribution matrix file (RMF) and the auxiliary response file (ARF)\footnote{RMF, ARF, and several ECF can be found at \webecf}. Within XSPEC, we load a spectral model by typing: \texttt{model tbabs*powerlaw}. A comprehensive list of spectral models can be found on the XSPEC webpage\footnote{\url{https://heasarc.gsfc.nasa.gov/xanadu/xspec/manual/Models.html}}. Every model will request to enter as input the initial parameters of the model. In our absorbed power law case, those parameters are the column density\footnote{In units of 10$^{22}$ cm$^{-2}$}, the photon index, and the normalization of the power law\footnote{The adopted value is not relevant for the ECF calculation.}. We produce a simulated spectrum using the command \texttt{fakeit none}. This spectrum will have the spectral shape previously defined by the user, in this example, an absorbed power law. At this stage, XSPEC will need the RMF and ARF files. XSPEC will also ask for the use of counting statistics in creating fake data, but since this is a simulated spectrum, this stage will not be needed. In the next steps requested by XSPEC (prefix definition, data file name, and exposure time setting) one can simply adopt the default values. Once every step is completed and the simulated spectrum is created, it is necessary to define the energy range of interest to calculate the corresponding ECFs. This is done by writing \texttt{ignore 0.0-0.2,2.3-**}. In this example, we are ignoring every spectral contribution between $0.0-0.2$ keV and everything larger than 2.3 keV. The command \texttt{show all} will print the ``model predicted rate'' of the spectrum on the screen. 
The model-predicted rate corresponds to the count rate of the model in the previously defined energy range. The command \texttt{flux 0.2 2.3} will return the flux of the spectral model at the desired energy range. The ratio between the count rate and the flux of the model gives the corresponding ECF value at the energy range of interest in units of cm$^{2}$ erg$^{-1}$. 

Table \ref{ecftable} summarizes three of the most common X-ray models and two sub-set where different parameters are given for each spectral model. We tabulate a multiplication factor that corresponds to the value needed to multiply the upper flux limit columns (UL\_B or UL\_S on the data product) to obtain a new upper limit that is based on and consistent with the tabulated spectral model. 

For our base-line spectral model, we obtain an ECF of $1.073\times10^{12}$ cm$^2$ erg$^{-1}$ in the $0.2-2.3$ keV energy range. We note that by modifying the photon index of the spectral model to $\Gamma=1.7$, the resulting ECF is $1.057\times10^{12}$ cm$^2$ erg$^{-1}$, approximately 4\% smaller than the ECF used in our upper limits.
We compute the ECF values for an absorbed black-body model (\texttt{tbabbs*bbody}), which is used to model spectra of e.g. neutron stars.
We compute the ECFs for two different black-body models: one has a black-body temperature of $kT=50\; \rm keV$, the other $kT=150\; \rm keV$. Both models use a galactic absorption of $N_{\rm H}=3\times10^{20} \; \rm cm^{-2}$.
The corresponding ECF are $7.75\times10^{11}$ 
cm$^2$ erg$^{-1}$ and $7.80\times10^{11}$ cm$^2$ erg$^{-1}$, respectively. The power law and black-body ECFs are calculated using abundances given by \cite{2000ApJ...542..914W} and cross-sections defined by \cite{1996ApJ...465..487V}. 

We also give ECFs for cool star templates, which are X-ray emitters due to their coronal emission. Cool stars tend to have moderately low intrinsic X-ray luminosities ($\log L_{\rm X} /[\rm erg\; s^{-1}]= 27 - 29$) unless they are very young. We, therefore, assume an optically thin thermal plasma model without an absorption component, since we expect cool stars of interest to be mainly located nearby and the ISM absorption in the eROSITA band is typically negligible within 150\,pc around the Sun. We use the XSPEC model \texttt{apec} with stellar coronal temperatures of 0.3 keV and 1.0 keV. We consider solar coronal abundances as given by \cite{1998SSRv...85..161G} to calculate the corresponding ECF in the $0.2-2.3$ keV energy band to be $1.202\times 10^{12}\,\mathrm{cm^2}\,\mathrm{erg^{-1}}$ and $1.212\times 10^{12}\,\mathrm{cm^2}\,\mathrm{erg^{-1}}$, respectively.

\begin{table}
\fontsize{7}{10}\selectfont
\caption{ECF table for the $0.2-2.3$ keV energy band.} 
\centering
\begin{center}
\begin{tabularx}{\columnwidth}{|X|X|X|X|}
\cline{3-4}
\hline
Model & Parameters & ECF\tablefootmark{a} (cm$^2$ erg$^{-1}$) & Multiplication Factor\tablefootmark{b} \\
\hline
\bfseries \multirow{4}{*}{\texttt{tbabs*pow}} &  \multirow{2}{8em}{\bm{$N_{\rm H}=3\times 10^{20}$}\textbf{cm}\bm{$^{-2}$} \bm{$\Gamma=2.0$}}  & \multirow{2}{8em}{\bm{$1.074\times10^{12}$} } &\bfseries  \multirow{2}{1em}{1} \\ 
&  &  & \\ 
\cline{2-4}
 & \multirow{2}{8em}{$N_{\rm H}=3\times 10^{20}\; \rm cm^{-2}$ $\Gamma=1.7$ }  & \multirow{2}{8em}{$1.060\times10^{12}$ } &\multirow{2}{1em}{1.013} \\ 
&  &  & \\ 
\hline
\multirow{4}{6em}{\texttt{tbabs*bbody}} & \multirow{2}{8em}{$N_{\rm H}=3\times 10^{20}\; \rm cm^{-2}$ $kT=0.05\; \rm keV$ }  & \multirow{2}{8em}{$7.566\times10^{11}$ } &\multirow{2}{1em}{1.419} \\ 
&  &  & \\ 
\cline{2-4}
& \multirow{2}{8em}{$N_{\rm H}=3\times 10^{20}\; \rm cm^{-2}$ $kT=0.15\; \rm keV$ }  & \multirow{2}{8em}{$1.229\times10^{12}$ } &\multirow{2}{1em}{0.874} \\ 
&  &  & \\ 
\hline
\multirow{2}{6em}{\texttt{apec}} & \multirow{2}{8em}{$kT=0.3\; \rm keV$ }  & \multirow{2}{8em}{$1.202\times10^{12}$ } &\multirow{2}{1em}{0.892} \\ 
&  &  & \\ 
\cline{2-4}
& \multirow{2}{8em}{$kT=1.0\; \rm keV$ }  & \multirow{2}{8em}{$1.212\times10^{12}$ } &\multirow{2}{1em}{0.885} \\ 
&  &  & \\ 
\hline
\end{tabularx}
\end{center}
\tablefoot{
\tablefootmark{a}{The values reported in the table have been approximated to the third significant digit. }
\tablefootmark{b}{The multiplication factor is defined as the correction value to multiply the upper flux limit column (UL\_B or UL\_S on the data product) to obtain the upper flux limit based on the tabulated spectral model. The small discrepancies regarding the multiplication factors and the listed ECFs are produced because the factors have been calculated using the full (non-approximated) ECF value. We highlight with bold letters the model and ECF used on our upper flux limit calculation. }}
\label{ecftable}
\end{table}

{We repeated the ECF calculation, now based on the sub-energy bands of the three-band detection.} The ECF values for the three-band and the entire energy band are given in Table \ref{ecftable3B}.

\begin{table*}
\fontsize{6}{10}\selectfont
\caption{ECF table for the soft ($0.2-0.6$ keV), medium ($0.6-2.3$ keV), hard ($2.3-5.0$ keV), and for the $0.2-5.0$ keV energy range.} 
\centering
\begin{center}
\begin{tabularx}{\textwidth}{|X|l|X|X|X|X|X|X|X|X|} 
\cline{3-10}
\multicolumn{1}{c}{}&\multicolumn{1}{c|}{}&\multicolumn{2}{c|}{0.2--0.6 keV}&\multicolumn{2}{c|}{0.6--2.3 keV}&\multicolumn{2}{c|}{2.3--5.0 keV}&\multicolumn{2}{c|}{0.2--5.0 keV}\\
\hline
Model & Parameters & \multicolumn{1}{c|}{ECF (cm$^2$ erg$^{-1}$)} &Multiplication Factor&  \multicolumn{1}{c|}{ECF (cm$^2$ erg$^{-1}$)}  &Multiplication Factor &  \multicolumn{1}{c|}{ECF (cm$^2$ erg$^{-1}$)} &Multiplication Factor &  \multicolumn{1}{c|}{ECF (cm$^2$ erg$^{-1}$)}  &Multiplication Factor \\
\hline
\bfseries \multirow{4}{1em}{\texttt{tbabs*pow}} &\multirow{2}{9em}{\bm{$N_{\rm H}=3\times 10^{20}$} \textbf{cm}\bm{$^{-2}$} \bm{$\Gamma=2.0$}}  & \multirow{2}{8em}{\bm{$1.026\times10^{12}$} } &\multirow{2}{1em}{\bm{$1$}}& \multirow{2}{8em}{\bm{$1.087\times10^{12}$} } &\multirow{2}{1em}{\bm{$1$}}& \multirow{2}{8em}{\bm{$1.147\times10^{11}$} } &\multirow{2}{1em}{\bm{$1$}}& \multirow{2}{8em}{\bm{$7.932\times10^{11}$} } &\multirow{2}{1em}{\bm{$1$}} \\ 
&  &  &&& &&&&\\ 
\cline{2-10}
 & \multirow{2}{10em}{$N_{\rm H}=3\times 10^{20}\; \rm cm^{-2}$ $\Gamma=1.7$ }  & \multirow{2}{8em}{$1.056\times10^{12}$ } &\multirow{2}{1em}{0.972}& \multirow{2}{8em}{$1.054\times10^{12}$ } &\multirow{2}{1em}{1.031}& \multirow{2}{8em}{$1.127\times10^{11}$ } &\multirow{2}{1em}{1.017}& \multirow{2}{8em}{$6.982\times10^{11}$ } &\multirow{2}{1em}{1.136}   \\ 
&  &  &&& &&&&\\ 
\hline
\multirow{4}{1em}{\texttt{tbabs*bbody}} & \multirow{2}{9em}{$N_{\rm H}=3\times 10^{20}\; \rm cm^{-2}$ $kT=50\; \rm eV$ }  & \multirow{2}{9em}{$7.492\times10^{11}$ } &\multirow{2}{1em}{1.370}& \multirow{2}{8em}{$1.239\times10^{12}$ } &\multirow{2}{1em}{0.877} & \multirow{2}{8em}{$3.063\times10^{11}$ } &\multirow{2}{1em}{0.374} & \multirow{2}{8em}{$7.626\times10^{11}$ } &\multirow{2}{1em}{1.040}   \\ 
&  &  &&& &&&&\\ 
\cline{2-10}
& \multirow{2}{9em}{$N_{\rm H}=3\times 10^{20}\; \rm cm^{-2}$ $kT=150\; \rm eV$ }  & \multirow{2}{8em}{$1.113\times10^{12}$ } &\multirow{2}{1em}{0.922} & \multirow{2}{8em}{$1.311\times10^{12}$ } &\multirow{2}{1em}{0.829}& \multirow{2}{8em}{$1.482\times10^{11}$ } &\multirow{2}{1em}{0.774} & \multirow{2}{8em}{$1.229\times10^{12}$ } &\multirow{2}{1em}{0.645}      \\ 
&  &  &&& &&&&\\ 
\hline
\multirow{2}{1em}{\texttt{apec}} & \multirow{2}{8em}{$kT=0.3\; \rm keV$ }  & \multirow{2}{8em}{$9.696\times10^{11}$ } &\multirow{2}{1em}{1.062}& \multirow{2}{8em}{$1.352\times10^{12}$ } &\multirow{2}{1em}{0.801}& \multirow{2}{8em}{$1.514\times10^{11}$ } &\multirow{2}{1em}{0.757} & \multirow{2}{8em}{$1.202\times10^{12}$ } &\multirow{2}{1em}{0.655}       \\ 
&  &  &&& &&&& \\ 
\cline{2-10}
& \multirow{2}{10em}{$kT=1.0\; \rm keV$ }  & \multirow{2}{8em}{$9.473\times10^{11}$ } &\multirow{2}{1em}{1.087} & \multirow{2}{8em}{$1.263\times10^{12}$ } &\multirow{2}{1em}{0.858} & \multirow{2}{8em}{$1.386\times10^{11}$ } &\multirow{2}{1em}{0.827}  & \multirow{2}{8em}{$1.147\times10^{12}$ } &\multirow{2}{1em}{0.686}   \\ 
&  &  &&& &&&&\\ 
\hline
\end{tabularx}
\end{center}
\tablefoot{Similarly to Table \ref{ecftable}, the values reported in the table have been approximated to the third significant digit. The small discrepancies regarding the multiplication factors and the listed ECFs are produced because the factors have been calculated based on the full (non-approximated) ECF value. We highlight with bold letters the model and ECF used on our upper flux limit calculation.}
\label{ecftable3B}
\end{table*}

\section{Retrieving the data from the tables}\label{appendix:retrievingul}
The name of our upper limit tables is of the form \texttt{em01\_174069\_024\_UpperLimitTable\_c010.fits}. In this example, the first letter, \texttt{e}, corresponds to the observation mode (\texttt{e}: survey or \texttt{s}: stacked catalogs), and the second letter, \texttt{m}, defines the ownership (\texttt{m}: eROSITA-DE, \texttt{i}: Russian sky, and \texttt{b}: shared sky area.). The number \texttt{01} corresponds to the first eROSITA all-sky scan, \texttt{174069} represents the sky tile number, and \texttt{024} is the eROSITA-internal coding of the energy band $0.2-2.3$ keV\footnote{The eROSITA internal codes for the energy bands at the soft ($0.2-0.6$ keV), medium ($0.6-2.3$ keV), hard ($2.3-5.0$ keV), and summed three-band ($0.2-5.0$ keV) are \texttt{021}, \texttt{022}, \texttt{023}, and \texttt{02e}, respectively}. 
The expression \texttt{c010} defines the software version used to produce the input data for the upper flux limit calculation.

Once the file is downloaded, in order to find the correct sky tile, one can make use of the file \texttt{SKYMAPS.fits}\footnote{Available at \url{https://erosita.mpe.mpg.de/dr1/AllSkySurveyData_dr1/}}, which lists the eROSITA sky tiles and their corresponding boundaries (ra$_{\rm min}$, dec$_{\rm min}$, ra$_{\rm max}$, and dec$_{\rm max}$). For a given position, the corresponding sky tile is identified by searching for $\rm ra_{min}< {\rm R.A.} < ra_{max}$ and $\rm dec_{min} < {\rm Dec.} < dec_{\max}$. \texttt{SKYMAPS.fits} also lists the owner of the sky tile (\texttt{e}, \texttt{i}, or \texttt{b}).

To compute the correct HEALPix index for a given set of coordinates, one can make use of the Python implementation of the HEALPix algorithm \texttt{astropy-healpix}, which has the necessary tasks to perform the coordinate transformations\footnote{\url{https://astropy-healpix.readthedocs.io/en/latest/coordinates.html#celestial-coordinates}}.
The HEALPix tesselation can be created as follows, \texttt{hp = HEALPix($\rm nside=2^{16}$, order=`nested', frame=`icrs’)}. This will create HEALPix cells with a resolution of $\sim$3\arcsec, similar to the eROSITA pixel size of 4\arcsec. 
The coordinate conversion can be done with the functions \texttt{skycoord\_to\_healpix(coords)} and \texttt{healpix\_to\_Skycoord(index)}, where \texttt{coords} is an array of \texttt{astropy.coordinates} and \texttt{index} is an array with HEALPix indexes. 
Finally, the corresponding upper limit can be retrieved by identifying the row of the computed HEALPix index (for the sky position of interest) in the downloaded table of the sky tile.

\end{appendix}

\bibliographystyle{aa}
\bibliography{bib.bib}

\end{document}